# General Modal Properties of Optical Resonances in Subwavelength Non-spherical Dielectric Structures

Lujun Huang[1], Yiling Yu[2], Linyou Cao[1, 2*]

[1]Department of Material Science and Engineering, [2]Department of Physics, North Carolina State University, Raleigh, NC 27695

**Abstract**

Subwavelength dielectric structures offer an attractive low-loss alternative to plasmonic structures for the development of resonant optics functionality such as metamaterials. Non-spherical like rectangular structures are of the most interest from the standpoint of device development due to fabrication convenience. However, no intuitive fundamental understanding of the optical resonance in non-spherical structures is available, which has substantially delayed the device development with dielectric materials. Here we elucidate the general fundamentals of optical resonances in non-spherical subwavelength dielectric structures of different shapes (rectangular or triangular) and dimensionalities (1D nanowires or 0D nanoparticles). We demonstrate that the optical properties (i.e. light absorption) of non-spherical structures are dictated by the eigenvalue of the structure's leaky modes. Leaky modes are defined as natural optical modes with propagating waves outside the structure. We also elucidate the dependence of the eigenvalue on physical features of the structure. The eigenvalue shows scaling invariance with the overall size, weakly relies on the refractive index, but linearly depends on the size ratio of different sides of the structure. We propose a modified Fabry-Perot model to account for this linear dependence. Knowledge of the dominant role of leaky modes and the dependence of the leaky mode on physical features can serve as a powerful guide for the rational design of devices with desired optical resonances. It opens up a pathway to design devices with functionality that has not been explored due to the lack of intuitive understanding, for instance, imaging devices able to sense incident angle, or superabsorbing photodetectors.

* To whom correspondence should be addressed.

Email: lcao2@ncsu.edu

The resonant light-matter interaction at subwavelength objects constitutes an important cornerstone for modern optics research. Much significance of the resonance has been manifested by the spectacular success of localized surface plasmon resonances in metallic nanostructures.[1-4] The plasmonic resonance, which results from the collective oscillation of free electrons, has enabled a plethora of exotic functionality, including negative refractive index,[5] superscattering,[6] electromagnetic cloaking,[7] extraordinary optical transmission,[8] plasmonic induced transparency,[9, 10] and beam control.[11] Significantly, recent studies have demonstrated that subwavelength dielectric structures can provide similarly strong, tunable resonances.[12, 13] The dielectric optical resonance provides an attractive low-loss alternative to plasmonic resonances due to the less lossy nature of dielectric materials (e.g. silicon) than metals (e.g. gold). It also offers a tantalizing prospect of monolithically integrating optical functionality into electric or optoelectronic devices that have overwhelmingly built on dielectric materials like silicon.

From the standpoint of device development, non-spherical such as rectangular dielectric structures are of the most interest. Rectangular structures, offering desirable fabrication convenience, can be massively fabricated with controlled physical features using the standard manufacturing processes in semiconductor industry. However, most of the current fundamental studies in dielectric resonances focus on circular cylinders or spherical particles, whose optical responses can be evaluated with analytical models such as Mie theory.[14, 15] While studies of these round structures provide useful insights, the rectangular structure may exhibit substantially different properties due to its faceted morphology. For very small structures with only the lowest resonant mode (i.e. dipole mode) involved, a non-spherical shape may not cause much difference in the optical resonance from spherical or circular shapes. This is because the field of dipole modes is very spread, typically extending far beyond the physical dimension of the structure, and hence not sensitive to fine morphological features.[16-18] But many important applications such as solar cells, biosensing, wavelength filtering, and Fano resonances would request using relatively big structures with involvement of higher modes that can provide better performance. The higher mode has greater field confinement and would be more sensitive to morphological features.[15] Substantial differences are expected between the spherical (or circular) and rectangular structures. Therefore, to better guide the device development on dielectric resonances, it is necessary to have an intuitive and quantitative understanding of the optical resonance in non-spherical like rectangular dielectric structures, which is however not available yet. For example, how would the optical resonance depend on the size or size ratio of rectangular structures?

Here we elucidate the general fundamentals of optical resonances in rectangular dielectric nanostructures. We demonstrate that the optical response (e.g. light absorption) of rectangular structures is dictated by the eigenvalue of leaky modes of the structure. Leaky mode is defined as a natural optical mode with propagating fields outside the structure.[19, 20] It is featured with a complex eigenvalue, whose real part dictates resonant conditions and imaginary part refers radiative loss.[19, 21] We also elucidate the dependence of the eigenvalue on physical features of the structure. The eigenvalue of leaky modes shows scale invariance with respect to the overall size of rectangular structures, and is weakly associated with the refractive index of the materials. Most significantly, the eigenvalue linearly depends on the size ratio $R$ of different sides of the structure. The linear dependence is related with the mode number $m$ and order number $l$ of leaky modes, and can be approximated as $(m-1)\pi R + (l-1)\pi$. We propose a modified Fabry-Perot model to reasonably account for the linear dependence. While focus is on one-dimensional (1D) nanowires (NWs), we also demonstrate similar dependence of the eigenvalue on the physical features in zero-dimensional (0D) nanoparticles (NPs) and structures with other shapes such as triangular.

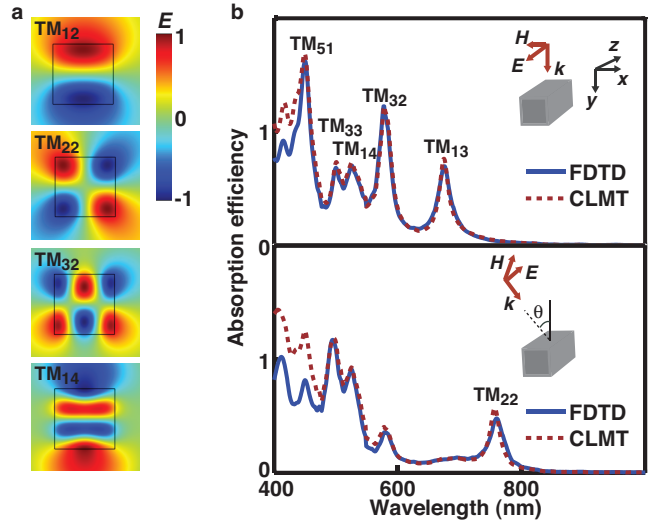

Figure 1. Role of leaky modes in the absorption of rectangular nanowires. (a) Calculated electric field distribution of typical TM-polarized leaky modes (electrical field is polarized with the nanowire axis). (b) Calculated absorption spectra of a square silicon nanowire in size of 200 nm illuminated by TM-polarized plane waves with incident angle of 0° (upper) and 45° (lower). The results calculated using the well-established FDTD technique (solid blue lines) and our CLMT model (dash red lines) are both presented. Some of the leaky modes involved in the absorption peaks are labeled as shown. Insets illustrate the illumination geometry with electric fields polarized parallel to the nanowire axis.

We start with examining the role of leaky modes in the optical response of rectangular dielectric nanostructures. Our previous studies have demonstrated that leaky modes dominate the optical response of spherical nanoparticles and circular nanowires, including light absorption and scattering.[12, 13, 19, 21] Here we use square silicon nanowires ($s$-SiNWs) as an example, and investigate the correlation between the absorption and the eigenvalue of the NW's leaky modes. Different from the leaky mode in spherical or circular structures, whose eigenvalue can be calculated using analytical models,[19, 22] no rigorous mathematic solutions are available for the leaky mode in rectangular structures. We calculate leaky modes in the square NW using finite element method (FEM). For simplicity, the refractive index of the materials is assumed to be a constant of 4 in the calculation. The calculated electric field distribution of typical leaky modes with transverse magnetic (TM) polarization (i.e. the electric field is parallel to the NW axis) is given in Figure 1a. Each of the leaky modes can be labeled using a mode number $m$ and an order number $l$ such as $TM_{ml}$. Physically, $m$ and $l$ are defined as the number of maxima in the electric field ($|E|^2$) distribution along the $x$ and $y$ axis of the square NW, respectively. The calculated eigenvalues of typical leaky modes are given in Table 1 (more results can be seen in Table S1-S2). The eigenvalue is a complex normalized parameter $nka$ ($nka = N_{real} - N_{imag}.i$). While the eigenvalue is calculated using a constant refractive index, it can reasonably apply to square NWs of arbitrary semiconductor materials. The real part of the eigenvalue $N_{real}$ dictates resonant conditions as optical resonances occur when the incident parameter $n_\lambda k_\lambda a$ matches

$$n_\lambda k_\lambda a = N_{real} \qquad (1)$$

where $n_\lambda$ is the real part of the refractive index of the materials at incident wavelength $\lambda$, $k_\lambda$ is the wavenumber of the incidence in free space ($\lambda = 2\pi/k_\lambda$), and $a$ is the size of the square NW. We also simulate the light absorption of the square SiNW with full field finite difference time domain (FDTD) techniques, in which the refractive index of silicon materials is used[23]. Figure 1b shows the simulated absorption spectrum of an *s*-SiNW in size of 200 nm illuminated by TM-polarized plane waves with different incident angles, 45° (oblique incidence) and 0° (normal incidence). By matching the eigenvalue of leaky modes (calculated using FEM) and the absorption spectra (calculated using FDTD), we can correlate each absorption peak to specific leaky modes (Fig. 1b). This suggests that the absorption peaks originate from resonances with the NW's leaky modes, as the spherical and circular structures that we studied previously.[12, 19]

Table 1. Eigenvalue of TM leaky modes in square nanowires ( n =4)

| $TM_{ml}$ | $l = 1$ | $l = 2$ | $l = 3$ | $l = 4$ |
|---|---|---|---|---|
| $m = 1$ | 1.51-0.61i | 4.07 – 0.30i | 7.10 - 0.10i | 9.91 - 0.31i |
| $m = 2$ | 4.07 – 0.30i | 6.17 - 0.096i | 8.66 - 0.15i | 11.5 - 0.0059i |
| $m = 3$ | 6.86 - 0.52i | 8.66 - 0.15i | 10.8 - 0.10i | 13.1 - 0.12i |
| $m = 4$ | 9.91 - 0.31i | 11.1 - 0.27i | 13.1 - 0.12i | 15.3 - 0.11i |

To more quantitatively understand the role of leaky modes, we use a model that we have previously developed,[19] coupled leaky mode theory (CLMT), to evaluate the light absorption of the square SiNW. The CLMT model considers the absorption of nanostructures as results from the coupling of incident light with leaky modes of the structure. It evaluates the light absorption using the eigenvalue of leaky modes, instead of by rigorously solving Maxwell equations with boundary conditions as all the existing techniques (i.e. FDTD and Mie theory) do. Assuming a single-mode nanostructure with a complex refractive index of $n$ ($n = n_{real} - n_{imag}.i$), its absorption efficiency (defined as the ratio of the absorption cross-section with respect to the geometrical cross-section) for an arbitrary incident wavelength $\lambda$ can be written as[19]

$$Q_{abs} = f(D) \frac{2/(q_{rad}q_{abs})}{4(\alpha-1)^2 + (1/q_{rad} + 1/q_{abs})^2} \qquad (2)$$

where $q_{abs}$ and $q_{rad}$ are the quality factors of the nanostructure due to absorption loss and radiative loss, respectively. $q_{abs}$ can be derived from the complex refractive index $n$, $q_{abs} = n_{real}/2n_{imag}$, and $q_{rad}$ is related with the eigenvalue of the leaky mode, $q_{rad} = N_{real}/2N_{imag}$. $\alpha$ indicates the offset of the incident wavelength $\lambda$ from the resonant wavelength $\lambda_0$ of the leaky mode as $\alpha = n_\lambda k_\lambda / n_0 k_0$, $n_0$ is the real part of the refractive index of the materials at $\lambda_0$ ($\lambda_0 = 2\pi/k_0$). $f(D)$ is the expansion coefficient of incident waves into characteristic harmonics of the coordinate system of the nanostructure. It can be analytically derived as $1/(k_\lambda r)$ for 1D circular NWs with $r$ the radius.[6, 19] However, there is no rigorous way to find out $f(D)$ for non-spherical structures. Without losing generality, we can set the $f(D)$ for square NWs as $C/(k_\lambda r)$ with $r = a/2$ and $C$ a expansion parameter that can be found out through numerical fitting (Table 2, the fitting process can be found in Supporting Information). Dictated by its origin as the expansion coefficient of incident waves in the nanostructure coordinate, $C$ varies with leaky modes, incident geometry, and the shape (i.e. size ratio) of the nanostructure, but is independent of the refractive index of the materials as well as the size of the NW. With the eigenvalue (Table 1) and the parameter $C$ (Table 2), we can use eq. (2) to calculate the absorption contributed by every single leaky mode in square NWs. The total absorption is a simple sum of the contribution from each individual mode.[19] The absorption spectra of the 200 nm-size $s$-SiNW calculated using eq. (2) is plotted in Figure 1b (dashed line). We can find that this calculation shows reasonable agreement with the result calculated using FDTD (solid line). This agreement further confirms the dominant role of leaky modes in the absorption of the square NW. It also indicates that eq. (2) provides a simple, intuitive, yet reasonably accurate approach for the evaluation of light absorption in non-spherical dielectric structures. A detailed list of the amplitude of $C$ for various leaky modes, incident angles, shapes of the nanostructure is given in Supporting Information. This list can be used as a database for the evaluation of light absorption in rectangular NWs of arbitrary semiconductor materials illuminated with arbitrary incident angles from eq. (2).

Table 2. Expansion parameter $C$ of plane waves for the leaky modes of square nanowires

| $TM_{ml}$ | Normal Incidence | | | | $\theta = 45°$ | | | |
|---|---|---|---|---|---|---|---|---|
| | $l=1$ | $l=2$ | $l=3$ | $l=4$ | $l=1$ | $l=2$ | $l=3$ | $l=4$ |
| $m=1$ | 1.00 | 1.80 | 1.75 | 1.75 | 1.0 | 0.9 | 0.05 | 1.1 |
| $m=2$ | 0 | 0 | 0 | 0 | 0.9 | 1.7 | 0.4 | 0.15 |
| $m=3$ | 1.00 | 3.05 | 1.32 | 2.65 | 1 | 0.4 | 0.7 | 0.5 |
| $m=4$ | 0 | 0 | 0 | 0 | 1.1 | 2.0 | 0.5 | 2.0 |

The correlation of optical properties with leaky modes can provide intuitive physical insights useful for device design. For instance, the imaginary part of the eigenvalue $N_{imag}$ indicates radiative loss.[19] Information of the $N_{imag}$ of leaky modes can guide choosing the right mode for specific applications. For light emission enhancement would ideally use the modes with low radiative loss (small $N_{imag}$), such as the $TM_{24}$ mode of square NWs, but for photodetectors should use the modes whose radiative loss matches the intrinsic absorption loss of the materials at the operation wavelength ($N_{real}/N_{imag} = n_{real}/n_{imag}$).[19] Additionally, we can find substantial differences in the absorption spectra under different incident angles (Fig. 1b). This is different from circular NWs, whose absorption is independent of the change in incident angle along the transvers direction due to circular symmetry. Our CLMT analysis indicates that this angle-dependent absorption is rooted in the expansion parameter $C$. Unlike circular NWs, in which $C$ is always unity,[6,19] the amplitude of $C$ for non-circular NWs is dependent on the incident angle (Table 2 and Fig. S4), which subsequently gives rise to angle dependence in the absorption. The amplitude of $C$ can be intuitively understood as the capability of incident light to excite specific leaky mode. We can find that none of the leaky modes with even mode number $m$ (such as $TM_{21}$, $TM_{22}$, $TM_{41}$, $TM_{42}$) can be excited by the normal incidence along the $y$ axis, as $C$ is zero for all these leaky modes. This is because the electric field distribution of these modes shows anti-symmetric with respect to the $y$ axis. In this case, the coupling of these modes with incident wave along the $y$ axis may be cancelled out due to the eigenfields with the same magnitude yet opposite parity. Knowledge of the angle-dependent $C$ may be helpful for the development of devices with angle-selective responses, for instance, photodetectors with capabilities to identify the angle of incident waves.

As the eigenvalue of leaky modes plays a dominant role in optical properties, it is important to better understand the fundamentals of the eigenvalue. For instance, how would the eigenvalue depend on the physical features (morphology, composition) of rectangular structures? Our studies indicate that the eigenvalue shows scale invariance with respect to the overall size of rectangular structures. Both the real part $N_{real}$ and imaginary part $N_{imag}$ of the eigenvalue do not change if each side of the structure is scaled by a common factor (Fig. 2a). According to eq. (1), this scale invariance means that the resonant wavelength of a given leaky mode would be linearly dependent on the size of the structure as $\lambda = M \cdot a$ with $M = 2\pi n_\lambda/N_{real}$. This is confirmed by the absorption spectra of square silicon NWs with different sizes (Fig. 2b). We can find that the wavelength of a given leaky mode (i.e. $TM_{13}$, $TM_{32}$, $TM_{33}$) indeed linearly increases with the size of the NW. Additionally, our calculation indicates that the eigenvalue is not very sensitive to the refractive index of the materials $n$. The real part of the eigenvalue $N_{real}$ is essentially independent of the refractive index, and the imaginary part $N_{imag}$ shows mild dependence (Fig. 3a). We can reasonably evaluate the absorption of square NWs of various semiconductor materials (Si, GaAs, CuInGaSe, etc.) from eq. (2) using the eigenvalue calculated with a constant refractive index of 4 (Fig. 3b-c), although these materials have different wavelength-dependent refractive indexes.[23] This relative insensitivity of the eigenvalue on the refractive index can substantially reduce the computation effort in the rational design of optical devices.

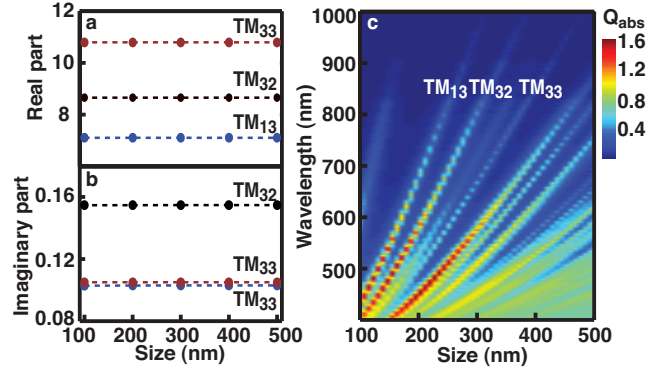

Figure 2. Scale invariance of the eigenvalue with respect to the size of dielectric structure. (a) The real part and (b) imaginary part of calculated eigenvalue for square NWs ($n = 4$) with different sizes. (c) The absorption efficiency $Q_{abs}$ of square silicon NWs plotted as a function of incident wavelengths (vertical axis) and size of the NW (horizontal axis). Typical leaky modes involved in the absorption are labeled as shown. We can see that generally the resonant wavelength linearly dependends on the size. The minor deviation from the linear dependence is caused by the wavelength dependence of the refractive index of silicon materials.

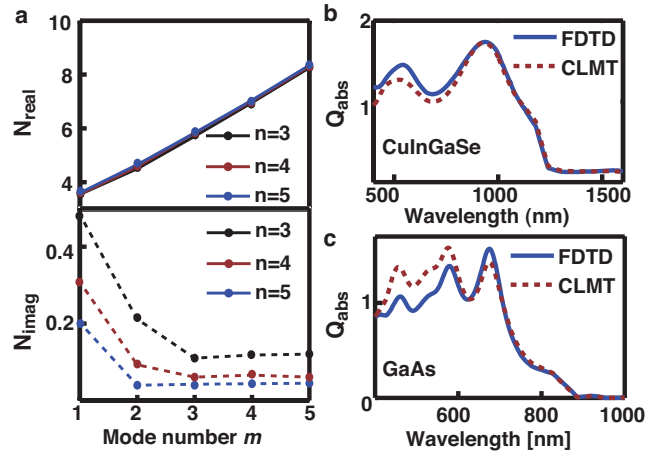

Figure 3. Insensitivity of the eigenvalue with respect to the refractive index of materials. (a) The real part (upper) and imaginary part (lower) of eigenvalues for leaky modes $TM_{m2}$ ($m = 1, 2, 3, 4, 5$) in square NWs calculated using different refractive indexes ($n = 3, 4,$ and $5$). (b-c) Calculated absorption spectra of 200 nm-size square NWs of CnInGaSe and GaAs for normal incidence of TM-polarized plane waves. The calculations using FDTD (solid blue lines) and our CLMT model (dash red lines) are both presented. The eigenvalues and expansion parameter $C$ listed in Table 1-2 were used in these CLMT calculations.

Very significantly, while the eigenvalue exhibits scale invariance with respect to the overall size of rectangular structures, it shows linear dependence on the size ratio of different sides of the structure. The linear dependence is closely related with the mode number $m$ and order number *l* of leaky modes. We use a rectangular NW with a width of *a* (along *x* axis) and a height of *b* (along *y* axis) as an example to illustrate this notion. Due to the size difference in the two sides, each leaky mode of the rectangular NW could have two equivalent eigenvalues depending on the specific side chosen as reference, *nka* (*a* is the reference side) or *nkb* (*b* is the reference side), for one eigen wavenumber *k*. Without losing generality, we define the size ratio *R* as $R = b/a$, and choose *b* as the reference side for the convenience of discussion, which gives the eigenvalue as *nkb* (the other eigenvalue *nka* can be correlated to *nkb* as $nka = nkb/R$). We can find that the real part $N_{real}$ of the eigenvalue shows linear dependence on the ratio *R* by a slope varying with the mode number *m* (Fig.4). The slope substantially increases with the mode number *m*, such as $TM_{11}$, $TM_{21}$, $TM_{31}$, and $TM_{41}$ (Fig.4a), but is more or less independent of the order number *l*, holding similar value for the modes with the same *m* yet different *l*, for instance, $TM_{11}$, $TM_{12}$, $TM_{13}$, and $TM_{14}$ (Fig.4b). To get more quantitatively understanding, we can fit the linear dependence as $nkb = s.R + t$, where *s* is the slope and *t* is the intercept with the vertical axis at $R = 0$. The fitting value of *s* and *t* for typical leaky modes are shown in Table 3. Interestingly, we find that *s* and *t* can be reasonably approximated as $s \approx (m-1)\pi$ and $t \approx (l-1)\pi$. It is worthwhile to note that, while the $N_{real}$ may be substantially dependent on the size ratio, the radiative quality factor ($q_{rad} = N_{real}/2N_{imag}$) generally shows much milder change with the size ratio (Figure S5). Similar linear dependence can also be found in TE leaky modes (Figure S6).

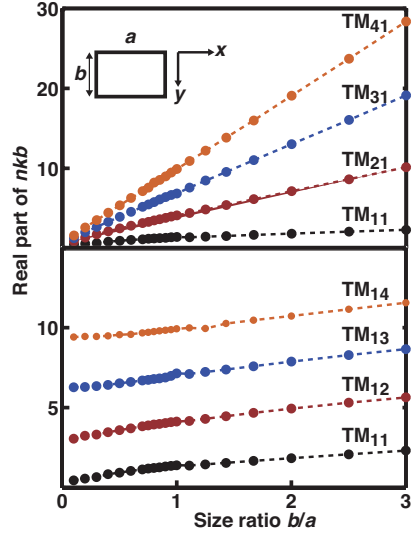

Figure 4. Linear dependence of the real part $N_{real}$ of the eigvenavlue $nkb$ on the size ratio of rectangular NWs $b/a$. Plotted are the calculated $N_{real}$ for (upper) leaky modes with the same $l$ yet different $m$ (upper) and (lower) leaky modes with the same $m$ yet different $l$ (lower). Inset is a schematic illustration for the rectangular NW.

Table 3. Fitted value of $s$ and $t$

|  | Slope $s$ ($\pi$) | | | | Intercept $t$ ($\pi$) | | | |
|---|---|---|---|---|---|---|---|---|
| $TM_{ml}$ | $l=1$ | $l=2$ | $l=3$ | $l=4$ | $l=1$ | $l=2$ | $l=3$ | $l=4$ |
| $m=1$ | 0.277 | 0.306 | 0.290 | 0.264 | 0.169 | 1.01 | 1.93 | 2.89 |
| $m=2$ | 1.03 | 1.00 | 0.923 | 0.901 | 0.283 | 0.975 | 1.84 | 2.71 |
| $m=3$ | 1.93 | 1.85 | 1.73 | 1.60 | 0.287 | 0.914 | 1.71 | 2.62 |
| $m=4$ | 2.86 | 2.66 | 2.54 | 2.45 | 0.271 | 0.904 | 1.65 | 2.41 |

The linear dependence of the eigenvalue on the size ratio can be accounted using a modified Fabry-Perot (FP) model. As shown in Figure 5a, the rectangular NW can be considered as a two-dimensional rectangular resonator with standing waves in both $x$ and $y$ axis. To start with, we assume the two standing waves are orthogonal to each other, in another word, no coupling between the FP resonances in the $x$ and $y$ axis. The wavevector $k$ of optical modes in the rectangular resonator can be written as $k^2 = k_x^2 + k_y^2$, where $k_x$ and $k_y$ are the wavevectors in the $x$ and $y$ axis, respectively. Correspondingly, the eigenvalue $nkb$ can be written as $(nkb)^2 =$ $(nk_xb)^2 + (nk_yb)^2 = (nk_xa)^2 \cdot R^2 + (nk_yb)^2$. When $R$ is infinitely small ($R \to 0$), the rectangular NW essentially evolves into a 1D Fabry-Perot resonator in the $y$ axis, and $nkb = nk_yb$. By linking this equation to $q$, the fitting value of $nkb$ at $R = 0$, we can derive the restriction on the eigenvector $k_y$ as $nk_yb = (l-1)\pi$. Similarly, we may have the restriction on $k_x$ as $nk_xa = (m-1)\pi$. As a result, the eigenvalue of an arbitrary leaky mode such as $TM_{ml}$ can be written as $nkb = \{[(m-1)\pi R]^2 + [(l-1)\pi]^2\}^{0.5}$. This simple equation indeed well matches the eigenvalue of high leaky modes with large $m$ and/or $l$ (for instance, $TM_{14,13}$) calculated using finite element method (FEM) (Fig. 5b, 5d). But it shows substantial deviation from the FEM calculation for low modes such as $TM_{22}$ (Fig. 5c-d). This is because the electromagnetic field of high modes is relatively more confined and thus the assumption that the FP resonances in the $x$ and $y$ axis do not couple with each other can reasonably apply. However, the FP resonances are expected to have strong coupling for low modes that are more leaky (more radiative loss). The coupling can be understood as resulting from the non-orthogonality of the fields in the $x$ and $y$ axis due to radiative loss at the interfaces. In order to reflect the coupling, we introduce a correction term $k_{corr}$ in the wavevector as $k^2 = k_x^2 + k_y^2 + k_{corr}^2$, where $k_{corr}^2$ is defined as $k_{corr}^2 = 2\beta k_x k_y$. The term of $k_x k_y$ in the correction is to intuitively reflect that the coupling involves the wavevectors in both $x$ and $y$ axis. $\beta$ is a phenomenon coupling coefficient. Its amplitude is determined by the integral overlap of the standing waves, and expected to drop to zero for high modes, in which the standing waves in the $x$ and $y$ axis are orthogonal. Therefore, the eigenvalue of low leaky modes can be approximately written as

$$nkb \approx \sqrt{(m-1)\pi^2 R^2 + (l-1)^2 \pi^2 + 2\beta(m-1)(l-1)\pi^2 R} \quad (3)$$

When the coupling coefficient $\alpha$ is close to be unity, eq. (3) can be simplified as $nkb \approx (m-1)\pi R + (l-1)\pi$. This indeed predicts a linear correlation between the eigenvalue ($N_{real}$) and the size ratio as shown in Fig.4. We can calculate the phenomenon coupling coefficient $\beta$ from eq. (3) as

$$\beta = \frac{(nkb)^2 - (m-1)\pi^2 R^2 - (l-1)^2\pi^2}{2(m-1)(l-1)\pi^2 R} \qquad (4)$$

The calculation result shows that the coupling coefficient decreases with $m$ or $l$ increasing and increases with the size ratio (Fig. 5e). This matches our expectation from intuitive perspective.

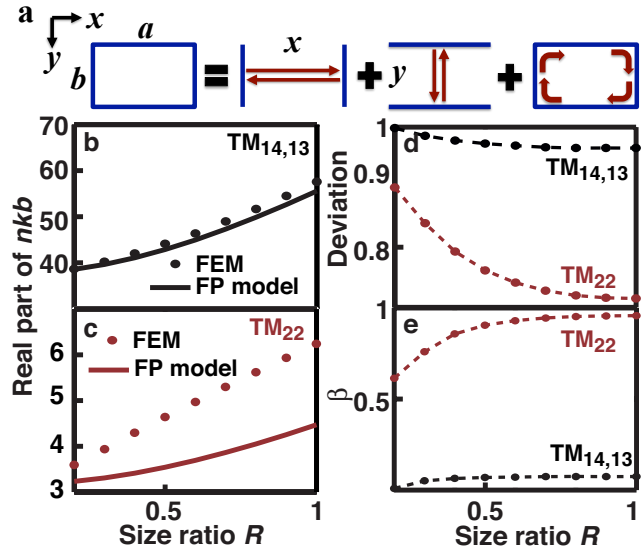

Figure 5. Modified Fabry-Perot resonator model for the leaky mode in rectangular dielectric NWs. (a) Schematic illustration of the model. The rectangular NW can be considered as a two-dimensional Fabry-Perot resonator with standing waves in both $x$ and $y$ axis coupling to each other. (b-c) The real part of the eigenvalue $nkb$ of a typical high mode $TM_{14,13}$ and a typical low mode $TM_{22}$ as a function of the size ratio $b/a$. The dot is the result calculated using finite element method FEM), and the solid line is calculated using the Fabry-Perot resonator model as $\{[(m-1)\pi R]^2 + [(l-1)\pi]^2\}^{0.5}$. (d) The ratio between the eigenvalues calculated using the FEM technique ($nkb$) and the FP model as $nkb/\{[(m-1)\pi R]^2 + [(l-1)\pi]^2\}^{0.5}$. (e) The phenomenon coupling coefficient $\beta$ of $TM_{14,13}$ and $TM_{22}$ as a function of the size ratio.

The linear dependence of $N_{real}$ on the size ratio provides great convenience for engineering the optical resonance of rectangular nanostuctures with arbitrary size ratios. In particular, the different correlation of the linear dependence with $m$ and $l$ offers capabilities to selectively tune the resonance of specific leaky modes. Without losing generality, we examine the evolution of resonant wavelengths of rectangular NWs by changing $b$ while fixing $a$ at 200 nm. We can readily predict the resonant wavelengths for an arbitrary $b$ from the linear dependence of the eigenvalue (Fig. 4) using eq. (1) as $\lambda = 2\pi nRa/N_{real}$. For simplicity, we assume the refractive index to be a constant of 4 ($n = 4$) in this analysis. The predicted resonant wavelengths for the modes with lower mode number $m$ ($TM_{12}$, $TM_{13}$, and $TM_{14}$) and higher $m$ ($TM_{31}$, $TM_{32}$, and $TM_{33}$) are plotted in Fig. 6a as a function of the size of $b$. We can see that the resonant wavelengths of low $m$ modes show drastic dependence on the size of $b$, quickly increasing with $b$. But the high $m$ modes generally show milder dependence. This different dependence of resonant wavelengths on the size $b$ can be confirmed by the absorption spectra of rectangular silicon NWs calculated using FDTD. We can find that the resonant wavelengths of $TM_{12}$, $TM_{13}$, and $TM_{14}$ linearly depend on the size of $b$ while the wavelength of $TM_{31}$ $TM_{32}$, and $TM_{33}$ shows milder non-linear dependence. Due to the structural symmetry, we can see similar dependence on the order number $l$ when fixing $b$ but changing $a$, in that case low order modes (such as $TM_{21}$, $TM_{31}$, and $TM_{41}$) show more drastic changes than the modes with high order number (such as $TM_{13}$, $TM_{23}$, and $TM_{33}$). This preferential dependence provides an unique capability to selectively engineer the resonant wavelengths of different modes. We can tune the resonant wavelengths of different leaky modes to be degenerate by controlling the size of $a$ or $b$. For instance, Figure 6b shows that the $TM_{12}$ and $TM_{31}$ modes of $s$-SiNWs can be tuned to have an identical resonant wavelength of ~ 600 nm at $b = 80$ nm. This engineering of mode degeneracy is necessary for the design of superscattering or superabsorbing structures. The scattering or absorption for a given incident wavelength has been demonstrated fundamentally limited by the number of modes that can resonate with the incidence.[6, 19] We can also see in Figure 6b that the absorption is indeed stronger wherever two modes are degenerate.

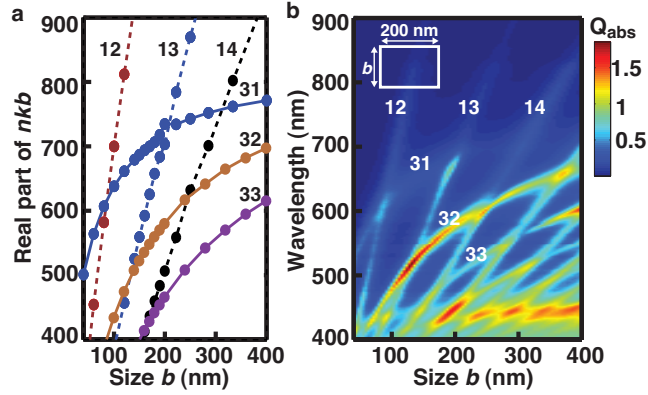

Figure 6. Selectively engineering of the resonances with leaky modes in rectangular NWs. (a) Resonant wavelengths of low $m$ modes ($TM_{12}$, $TM_{13}$, and $TM_{14}$) and high $m$ modes ($TM_{31}$, $TM_{32}$, and $TM_{33}$) as a function of the size $b$ with $a$ fixed at 200 nm. These resonant wavelengths are derived from the linear dependence of $N_{real}$ shown in Figure 4. (b) Calculated (using FDTD) absorption efficiency of rectangular SiNWs as a function of the size $b$ ($a$ fixed at 200 nm) and incident wavelengths. The leaky modes associated with resonant absorption peaks are given in the figure. For visual convenience, the leaky modes are only labeled with the subscript numbers.

The linear dependence of the eigenvalue of leaky modes on size ratio generally exists in subwavelength non-spherical dielectric structures. Except the rectangular NWs, we also find similar linear dependence in zero-dimensional (0D) rectangular particles and structures with other shapes such as triangular. The leaky mode in rectangular particles can be labeled using three subscript numbers $m, l, j$, which are defined as the number of maxima in the field distribution along $x, y$ and $z$ axis, respectively. It is worthwhile to note that, different from 1D NWs that the polarization of leaky modes can be distinguished as TE or TM, we find that there are no well defined TE or TM modes in rectangular nanoparticles. Without losing generality, for a particle with dimension of $a, b, c$ as illustrated in Fig. 7 inset, we define the eigenvalue as $nkc$ ($c$ is chosen as the reference side) and two size ratios $R_1 = c/a$ and $R_2 = c/b$, and examine the eigenvalue as a function of the size ratios. Our analysis indicates that the real part of the eigenvalue can be approximated as $nkc \approx (m-1)\pi R_1 + (l-1)\pi R_2 + (j-1)\pi$. For example, Figure 7 shows the eigenvalue of typical leaky modes in rectangular particles as a function of the size ratio $R_1$. In this calculation $a$ and $b$ are arbitrarily set to be a fixed ratio of 2, $a/b = 2$. The eigenvalue can be seen linearly dependent on the size ratio, and the linear dependence can be fitted as $nkc = s \cdot R_1 + t$. The fitted value of $s$ and $t$ (see Table S9 in the Supporting Information) can be approximated as $s \approx (m-1)\pi + 2(l-1)\pi$ and $t \approx (j-1)\pi$. This linear dependence can also be accounted by the modified FP model in which the coupling between the standing waves in $x$, $y$, and $z$ axis are taken into account (see Supporting Information Table S7). Additionally, we find that he eigenvalue of triangular NWs is linearly dependent on the size ratio of the height and the bottom of the triangle (see Supporting Information Fig. S7-S8).

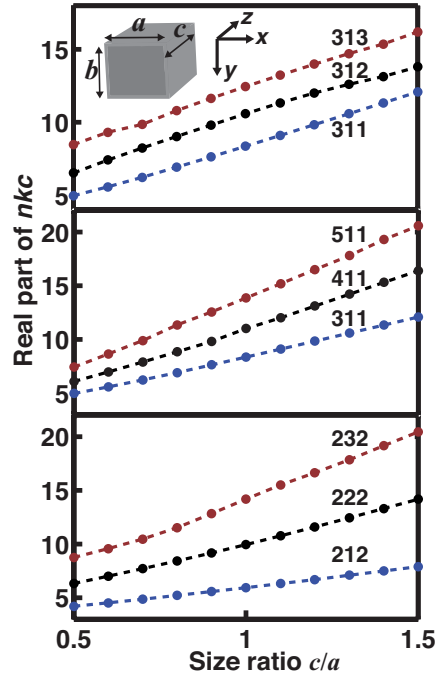

Figure 7. Linear dependence of the real part of the eigenvalue $nkc$ on the size ratio $c/a$ of rectangular particles. Each of the modes is labeled by three subscript number, $m$, $l$, and $j$. Plotted are the calculated $N_{real}$ for (upper) leaky modes with the same $m$ and $l$ yet different $j$, (center) leaky modes with the same $l$ and $j$ yet different $m$, and (lower) leaky modes with the same $m$ and $j$ yet different $l$. Inset is a schematic illustration of the rectangular particle.

In conclusion, we elucidate the general fundamentals of optical resonances in non-spherical dielectric structures, including different dimensionalities (1D nanowires and 0D nanoparticles) and different shapes (rectangular and triangular). We demonstrate that the resonant optical responses of non-spherical dielectric structures are dictated by the eigenvalue of leaky modes. We also discover the dependence of the eigenvalue on physical features (size and size ratio, composition) of the non-spherical structure. The eigenvalue shows scale invariance with respect to the overall size of the structure, is weakly associated with the refractive index of the materials, and is linearly dependent on the size ratio of different sides of the structure. Knowledge of the dominant role of leaky modes and the dependence of the leaky mode on physical features can serve as a powerful guide for the rational design of optical devices with desired optical resonances. For instance, the scale invariance indicates that the optical resonance can be readily tuned to be at arbitrary incident wavelength by control of the size. Information of the angle-dependent expansion parameter can provide guidance for the development of devices with angle-selective optical response, such as sensing the angle of incident light. The preferential linear dependence of the eigenvalue offers a useful capability to engineer the degeneracy of leaky modes, which is important for the development of superscattering or superabsorbing devices.

## Acknowledgements

This work is supported by start-up fund from North Carolina State University. L. Cao acknowledges a Ralph E. Power Junior Faculty Enhancement Award for Oak Ridge Associated Universities.